\documentclass[aps,prx,reprint]{revtex4-1}
\usepackage{blindtext}
\usepackage{graphicx}
\usepackage{amsmath}
\usepackage{amssymb}
\usepackage[english]{babel}
\usepackage{xcolor}
\usepackage{siunitx}
\usepackage[colorlinks=true, pdfstartview=FitV, linkcolor=blue,
citecolor=blue, urlcolor=blue]{hyperref}
\usepackage[capitalise]{cleveref}
\usepackage{float}

\makeatletter
\let\saved@includegraphics\includegraphics
\AtBeginDocument{\let\includegraphics\saved@includegraphics}
\renewenvironment*{figure}{\@float{figure}}{\end@float}

\makeatother

\makeatletter
\newcommand\footnoteref[1]{\protected@xdef\@thefnmark{\ref{#1}}\@footnotemark}
\makeatother

\begin{document}
\title{Fully tunable hyperfine interactions of hole spin qubits in Si and Ge quantum dots}

\author{Stefano Bosco}
\email{stefano.bosco@unibas.ch}
\affiliation{Department of Physics, University of Basel, Klingelbergstrasse 82, 4056 Basel, Switzerland}
\author{Daniel Loss}
\affiliation{Department of Physics, University of Basel, Klingelbergstrasse 82, 4056 Basel, Switzerland}

\begin{abstract}
Hole spin qubits are frontrunner platforms for scalable quantum computers, but state-of-the-art devices suffer from noise originating from the hyperfine interactions with nuclear defects. 
We show that these interactions have a highly tunable anisotropy that is controlled by device design and external electric fields. This tunability enables sweet spots where the hyperfine noise is suppressed by an order of magnitude and is comparable to isotopically  purified materials. We identify surprisingly simple designs where the qubits are highly coherent and are largely unaffected by both charge and hyperfine noise.
We find that the large spin-orbit interaction typical of elongated quantum dots not only speeds up qubit operations, but also dramatically renormalizes the hyperfine noise, altering qualitatively the dynamics of driven qubits and enhancing the fidelity of qubit gates.
Our findings serve as guidelines to design high performance qubits for scaling up quantum computers.
\end{abstract}

\maketitle

\paragraph*{Introduction.}
Spin qubits in hole quantum dots are leading candidates to process quantum information~\cite{scappucci2020germanium,doi:10.1146/annurev-conmatphys-030212-184248,hendrickx2020four}.
Elongated hole quantum dots hold particular promise because of their large and tunable direct Rashba spin-orbit interaction (DRSOI)~\cite{DRkloeffel1,DRkloeffel2,DRkloeffel3,bosco2021squeezed}. This property enables fast all-electric gates~\cite{maurand2016cmos,watzinger2018germanium,camenzind2021spin,froning2020ultrafast,wang2020ultrafast}  via electric dipole spin resonance (EDSR)~\cite{PhysRevB.74.165319,PhysRevLett.98.097202} and could pave the way towards scalable quantum computers.
A major source of decoherence in these qubits comes from the hyperfine interaction between the confined hole and the nuclear spins.
Isotopic purification could decrease the number of nuclear defects by two~\cite{veldhorst2014addressable,Yonedaquantumdotspinqubit2018} to three~\cite{tyryshkin2012electron} orders of magnitude, however this procedure is expensive and not routine  in state-of-the-art devices, where hyperfine noise still drastically limits the qubit performance~\cite{camenzind2021spin}.

The hyperfine interactions in hole nanostructures originate from dipolar coupling~\cite{warburton2013single,PhysRevB.100.085305,PhysRevB.79.195440,PhysRevB.78.155329,PhysRevLett.102.146601} and in Si and Ge their amplitude is predicted to be comparable to electrons~\cite{PhysRevB.101.115302,philippopoulos2020hyperfine}. In contrast to electrons~\cite{PhysRevB.65.205309,PhysRevLett.88.186802,PhysRevB.70.195340}, however, these interactions are strongly anisotropic, approaching the Ising limit in planar structures~\cite{PhysRevB.78.155329,belhadj2010impact}. Ising coupling causes a slow power-law spin decay depending on the orientation of the field~\cite{PhysRevLett.99.106803,PhysRevB.72.125337} and could result in an enhancement of the qubit lifetime when the nuclear spins are suitably prepared~\cite{PhysRevLett.105.266603, PhysRevB.85.195323,PhysRevB.73.205302,prechtel2016decoupling}.

In this work, we show that in elongated hole quantum dots the amplitude of the hyperfine interactions as well as their anisotropy are fully tunable by external electric fields, resulting in devices with an order of magnitude smaller hyperfine noise. This decrease in noise can save two orders of magnitude of isotopic purification.  Strikingly, by analyzing common designs, we find optimal working points in Si Fin field effect transistors (FinFETs)~\cite{camenzind2021spin,geyer2020silicon,kuhlmann2018ambipolar} where both hyperfine and charge noise~\cite{bosco2020hole} are suppressed simultaneously, strongly boosting the qubit coherence.
Also, we examine the interplay between DRSOI and hyperfine interactions~\cite{PhysRevB.90.245305}. The large DRSOI substantially renormalizes the hyperfine noise and qualitatively alters the spin dynamics in EDSR experiments, yielding more coherent qubit operations. 
We foresee that the tunability of the hyperfine interactions could be exploited also in hybrid systems~\cite{Hensensiliconquantumdotcouplednuclear2020} to engineer a coherent coupling between quantum dots and nuclear spin qubits.

\paragraph*{Hyperfine interactions in elongated quantum dots.}

Spin qubits confined in quantum dots are generally described by the effective Hamiltonian~\cite{PhysRevB.78.155329,PhysRevB.100.085305,PhysRevB.79.195440} $H_\text{Q}=(\pmb{b}+\pmb{h})\cdot\pmb{\sigma}/2$, comprising a Zeeman field $\pmb{b}=\mu_B\underline{g}\textbf{B}$ and an Overhauser field $\pmb{h}=\sum_k {A_k}\Gamma(\textbf{r}_k)\textbf{I}^k/n_0$.
These fields model the magnetic interactions of the confined particle with an external magnetic field $\textbf{B}$ and with an ensemble of nuclear spins $\textbf{I}^k$ at position $\textbf{r}_k$, respectively. Here, $\mu_B$ is the Bohr magneton, $n_0$ is the nuclear density, $A_k$ is the hyperfine coupling strength, and we neglect  small long-range corrections to the hyperfine interactions~\cite{PhysRevB.78.155329}.

The microscopic properties of the system determine the values of the matrices $\underline{g}$ of $g$ factors and $\Gamma(\textbf{r}_k)$ of local spin susceptibilities.
While in electron quantum dots these matrices are proportional to the identity matrix, resulting in isotropic interactions, in hole dots they have a richer structure and heavily depend on the mixing of heavy (HH) and light hole (LH) bands~\cite{WinklerSpinOrbitCoupling2003}, that carry spin $3/2$ and $1/2$, respectively.
In particular, HH dots have a strongly anisotropic $g$-factor~\cite{watzinger2016heavy} and Ising hyperfine interactions $\propto \sigma_z h_z$~\cite{PhysRevB.78.155329,PhysRevB.100.085305,PhysRevB.79.195440}, while in LH dots the anisotropy is less pronounced and the transverse components $h_{x,y}$ of $\pmb{h}$ are two times larger than $h_z$~\cite{PhysRevB.101.115302}.

In this work, we analyze quantum dots that are tightly confined in the $\pmb{\rho}=(x,y)$ plane and extend in the $z$-direction.
In these systems, the HH-LH mixing can be engineered by designing the dot~\cite{DRkloeffel3,bosco2020hole} and is highly tunable by external electric fields. It is accurately modelled by the Hamiltonian
\begin{equation}
\label{eq:total-H-3D}
H= H_\text{LK}+V_\text{C}(\pmb{\rho})+\frac{\hbar\omega_z}{2l_z^2} z^2-eE_y y \ ,
\end{equation}
that includes the Luttinger-Kohn Hamiltonian $H_\text{LK}$~\cite{WinklerSpinOrbitCoupling2003} and an electric field $E_y$ perpendicular to the long direction. Here, the dot is defined by an harmonic potential with frequency $\omega_z$ and length $l_z$, and an abrupt potential $V_\text{C}(\pmb{\rho})$. This potential models etched nanowires, but we emphasize that our theory also describes squeezed dots in planar heterostructures~\cite{bosco2021squeezed}. 

When the nanowires are grown along high symmetry axes, e.g. $z\parallel[001]$ or $[110]$, the ground state Kramer partners $\Psi_{\uparrow\downarrow}$ of $H$ are well-approximated by $\Psi_{\uparrow\downarrow}\approx e^{-i z \sigma_x/l_{SO}} \psi_{\uparrow\downarrow}(\textbf{r})$~\cite{SM-hyper}.  The spinors
\begin{equation}
\label{eq:WF}
\psi_\uparrow(\textbf{r}) =\varphi(z)\left(\psi_H(\pmb{\rho})\left|+\frac{3}{2}\right\rangle+ \psi_L(\pmb{\rho})\left|-\frac{1}{2}\right\rangle\right)=\mathcal{T}\psi_\downarrow(\textbf{r}) \ 
\end{equation}
locally hybridize different eigenstates of the spin-3/2 matrix $J_z$ by the spin-resolved and $E_y$-dependent wavefunctions $\psi_{H,L}(\pmb{\rho})$.
The time-reversal operator $\mathcal{T}$ flips the spin and complex conjugates the functions, and  $\varphi(z)$ is the harmonic oscillator ground state.
The  spin-dependent local phase in $\Psi_{\uparrow\downarrow}$ accounts for the large DRSOI~\cite{DRkloeffel1,DRkloeffel2,DRkloeffel3,bosco2021squeezed}, which is parametrized by a $E_y$-dependent spin-orbit length $l_{SO}$ typically of tens of nanometers~\cite{camenzind2021spin,froning2020ultrafast}. Here, we compute $\psi_{H,L}$ and $l_{SO}$ by numerically discretizing Eq.~\eqref{eq:total-H-3D} in analogy to~\cite{bosco2020hole}. 
Because we only study dots where $E_y$ is aligned to high symmetry axes, the DRSOI points in the $x$-direction~\cite{DRkloeffel1,DRkloeffel2,DRkloeffel3,bosco2020hole}; more general cases, also including strain and high energy hole bands, are discussed in~\cite{SM-hyper}.

When $|\textbf{B}|\lesssim 1$~T, the magnetic interactions are weaker than $\hbar\omega_z$ and by projecting the hyperfine $H_\text{HF}= \sum_{k} \delta(\textbf{r}-\textbf{r}_k) A_k\textbf{I}^k\cdot \textbf{J}/2n_0$~\cite{PhysRevB.78.155329,PhysRevB.101.115302,PhysRevB.100.085305,PhysRevB.79.195440,PhysRevB.65.205309} and Zeeman Hamiltonian $H_\text{Z}=2  \kappa \mu_B  \textbf{B}\cdot \textbf{J}$~\cite{WinklerSpinOrbitCoupling2003} onto $\Psi_{\uparrow\downarrow}$, we find~\cite{SM-hyper} 
\begin{equation}
\label{eq:gamma}
\Gamma(\textbf{r})=|\varphi(z)|^2R_x\left(\frac{2z}{l_{SO}}\right)\left(
\begin{array}{ccc}
  \text{Re}[\gamma_+ ]  &  \text{Im}[\gamma_- ] & 0  \\
  -\text{Im}[\gamma_+ ]  & \text{Re}[\gamma_- ]   & 0 \\
 0 & 0 & \gamma_z  \\
\end{array}
\right) \ ,
\end{equation}
and $g_{ij}=4\kappa \int d\textbf{r}\Gamma_{ij}(\textbf{r})$.
Here, we neglect small corrections coming from terms~$\propto J^3_i$~\cite{WinklerSpinOrbitCoupling2003,PhysRevB.101.115302}, $\propto 1/l_z^2$~\cite{DRkloeffel2,bosco2021squeezed,adelsbergerorbital},  and from magnetic orbital effects~\cite{PhysRevResearch.3.013081,adelsbergerorbital}.
We define  $\gamma_\pm=\psi_L(\psi_L\pm\sqrt{3}\psi_H)$, $\gamma_z=(3|\psi_H|^2-|\psi_L|^2)/2$,  and $R_x(\theta)$ is an anticlockwise rotation matrix of an angle $\theta$ around $x$. This SOI-dependent rotation causes the well-known suppression $e^{-l_z^2/l_{SO}^2}$ of the $g$-factor~\cite{PhysRevB.77.045434,PhysRevResearch.3.013081}, and also significantly alters the hyperfine interactions.

If no effort is put in preparing the state of the nuclear spins~\cite{PhysRevLett.105.266603,PhysRevB.85.195323,PhysRevB.73.205302, prechtel2016decoupling}, the Overhauser field $\pmb{h}$ is Gaussian distributed~\cite{PhysRevLett.88.186802,PhysRevB.78.195302,PhysRevB.65.205309,PhysRevB.72.125337}, and has zero mean and diagonal covariance matrix $\Sigma_{ij}= {\hbar^2\sigma_{i}} \delta_{ij}/{\bar{\tau}^2}$~\cite{SM-hyper}. The characteristic time of hyperfine-induced qubit decay is
\begin{equation}
\label{eq:time_tau}
\bar{\tau}=\frac{\hbar}{|A_k|}\sqrt{\frac{3 N}{\nu I(I+1)}} \ ,
\end{equation}
where  $\nu$ is the isotopic abundance of the nuclear defects, $N=\sqrt{2\pi}l_z \mathcal{A}_{\rho}n_0$ is the number of atoms in the dot, and $\mathcal{A}_{\rho}$ is the area of the wire cross-section. In particular, for natural Si and Ge dots with  $N\approx 10^4$ atoms~\cite{geyer2020silicon,kuhlmann2018ambipolar,camenzind2021spin},
we find $\bar{\tau}_\text{Si}\approx 0.36$~$\mu$s and $\bar{\tau}_\text{Ge}\approx 0.11$~$\mu$s~\footnote{
We use $\nu_\text{Si}=4.7 \%$, $I_\text{Si}=1/2$, $|A_k^\text{Si}|=1.67$~$\mu$eV, and $\nu_\text{Ge}=7.7 \%$, $I_\text{Ge}=9/2$, $|A_k^\text{Ge}|=0.73$~$\mu$eV;  $A_k^i$ is related  by   $A_k^i\equiv 2A_\parallel^i/3$ to the quantity $A_\parallel^i$ estimated in~\cite{PhysRevB.101.115302} and~\cite{philippopoulos2020hyperfine} for Si and Ge, respectively.
}.
The dimensionless diagonal elements of $\Sigma$ are
\begin{equation}
\label{eq:sigma_SOI}
\sigma_x=\sigma_x^0 \ , \ \text{and}  \ \sigma_{y,z}= \sigma_{M}+e^{-\frac{2l_z^2}{l_{SO}^2}}(\sigma_{y,z}^0-\sigma_{M})  \ ,
\end{equation}
and at $l_{SO}^{-1}=0$ they attain the values
\begin{subequations}
\label{eq:integrals}
\begin{align}
\sigma_{x,y}^0 &=\mathcal{A}_{\rho}\!\int\!d\pmb{\rho}\left( |\psi_L|^4+3|\psi_L\psi_H|^2 \pm 2\sqrt{3}\text{Re}\!\left[\psi_L^3\psi_H\right]\right) \ , \\
\sigma_{z}^0 &= \mathcal{A}_{\rho}\!\int\!d\pmb{\rho}\left(\frac{3}{2}|\psi_H|^2-\frac{1}{2}|\psi_L|^2\right)^2 \ .
\end{align}
\end{subequations}
 The DRSOI renormalizes $\sigma_{y,z}$ to the mean value $\sigma_{M}= (\sigma_{y}^0+\sigma_{z}^0)/2$: this renormalization has important consequences on the qubit dynamics.

\paragraph*{Tunable hyperfine noise in spin qubits.}

\begin{figure}[t]
\centering
\includegraphics[width=0.5\textwidth]{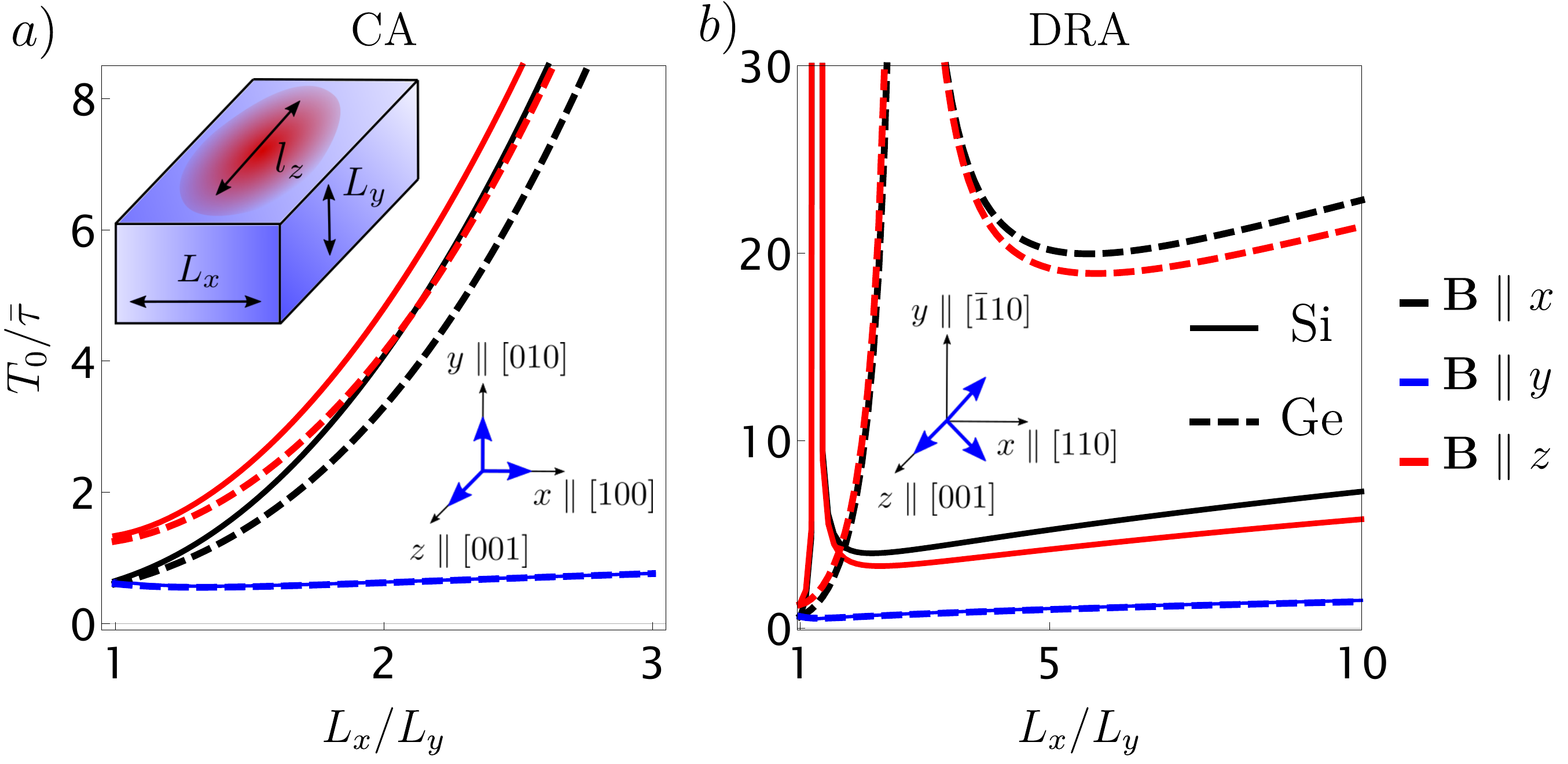}
\caption{\label{fig:Ising} Gaussian decay time $T_0$ in idle qubits in rectangular wires. We examine Si (solid lines) and Ge (dashed lines) dots when $E_y=0$ and show $T_0$ for different directions of $\textbf{B}$ as a function of  $L_x/L_y$.  In a) and b), we study wires grown along the CA and DRA, respectively. Confinement (crystallographic) axes are shown with black (blue) arrows. }
\end{figure}

Because of the anisotropy of the hyperfine interactions, the spin dynamics strongly depends on the direction of the external fields. 
For simplicity, we consider $\textbf{B}$ aligned along the confinement axes, such that $\pmb{b}\parallel \textbf{B}$~\cite{PhysRevResearch.2.033036, bellentani2021towards}, and we analyze the dephasing of an idle qubit by finding the transition probability $P(t)=\langle|\langle-|e^{-iH_Qt/\hbar}|+\rangle|^2\rangle_g$ between the states $|\pm\rangle=(|0\rangle\pm|1\rangle)/\sqrt{2}$.
Here, $|0,1\rangle$ are eigenstates of $\pmb{b}\cdot\pmb{\sigma}$ and the outer brackets  indicate the average of $\pmb{h}$ over a Gaussian distribution with covariance~${\Sigma}$.

For typical values $|\textbf{B}|\sim 100$~mT, the hyperfine broadening is small, i.e. $|\pmb{b}|\bar{\tau}/\hbar\sqrt{\sigma_i}\gg 1$, and
\begin{equation}
\label{eq:P_idle}
P(t)\approx \frac{1}{2}-\frac{1}{2}\frac{e^{-\frac{t^2}{2T_0^2}}\cos[\omega_B t+ \phi(t)/2]}{\sqrt[4]{(1+t^2/\tau_1^2)(1+t^2/\tau_2^2)}}  \ , 
\end{equation}
where $\phi(t)=\arctan(t/\tau_1)+\arctan(t/\tau_2)$. 
The hyperfine interactions parallel to $\pmb{b}$ dampen the coherent spin precession with frequency $\omega_B=|\pmb{b}|/\hbar$ by a Gaussian factor with time scale $T_0=\bar{\tau}/\sqrt{\sigma_\parallel}$, while the transverse interactions cause a power law decay with time scales $\tau_{i}=\omega_B\bar{\tau}^2/\sigma_\perp^i $~\cite{PhysRevB.78.155329,PhysRevB.79.195440}. We call $\sigma_\parallel$ and $\sigma_\perp^{1,2}$ the dimensionless diagonal elements of ${\Sigma}$ in Eq.~\eqref{eq:sigma_SOI} parallel and perpendicular to $\pmb{b}$, respectively.
Eq.~\eqref{eq:P_idle} is derived in~\cite{SM-hyper}, including also arbitrary field directions.

The power law tail is observable when $\tau_{1,2}/T_0=\omega_B\bar{\tau}\sqrt{\sigma_\parallel}/\sigma_\perp^{1,2}\lesssim 1$, a condition that requires highly anisotropic hyperfine interactions when $\omega_B$ is in the GHz range.
This anisotropy can be engineered by the confinement potential. For example, in rectangular wires~\cite{maurand2016cmos} grown along the crystallographic axes (CA), high aspect ratios $L_x/L_y\gg 1$ enable Ising hyperfine interactions $\propto h_y\sigma_y$ because the groundstate comprises HH polarized along the tighter confinement direction~\cite{PhysRevB.78.155329,PhysRevB.101.115302,PhysRevB.100.085305,PhysRevB.79.195440}.
In Fig.~\ref{fig:Ising}a), we show that the anisotropy decreases in typical Si and Ge wires where $L_x\sim L_y$, resulting in a fast Gaussian qubit decay for any direction of $\textbf{B}$. In particular, at $L_x=L_y$ the Gaussian  times $T_0^{i}$  at $\textbf{B}\parallel i=\{x,y,z\}$ are related by $T_0^x=T_0^y\approx T_0^z/2$, consistent with LH  dots~\cite{PhysRevB.101.115302}.

The hyperfine interactions are strikingly different when the cross-section is rotated by $\pi/4$ with respect to $z$, as illustrated in Fig.~\ref{fig:Ising}b). We call this orientation direct Rashba axes (DRA)~\cite{bosco2020hole} because it guarantees the largest DRSOI in wires~\cite{DRkloeffel3,bosco2021squeezed,PhysRevB.99.115317}. 
First, because of the sizeable HH-LH mixing even in planar heterostructures, yielding DRSOI~\cite{PhysRevB.103.085309}, the hyperfine interactions are non-Ising at $L_x/L_y\gg 1$. The hyperfine anisotropy is still pronounced in wide Ge wires, but it decreases notably in Si, where the spin decay remains Gaussian with times $T_0^i$ of hundreds of nanoseconds.
Surprisingly, however, we find that Ising interactions are recovered at specific aspect ratios $L_x/L_y\approx 1.3\ (2.7)$ in Si (Ge). At these points  $T_0\rightarrow\infty$ when $\textbf{B}\perp y$, see Fig.~\ref{fig:Ising}b), resulting in sweet spots where the  qubit lifetime is largely enhanced and where the spin decay has a slow power-law tail with a longer timescale $\tau_1=\omega_B\bar{\tau}^2/\sigma_y$ of tens of microseconds. \\

\begin{figure}[t]
\centering
\includegraphics[width=0.5\textwidth]{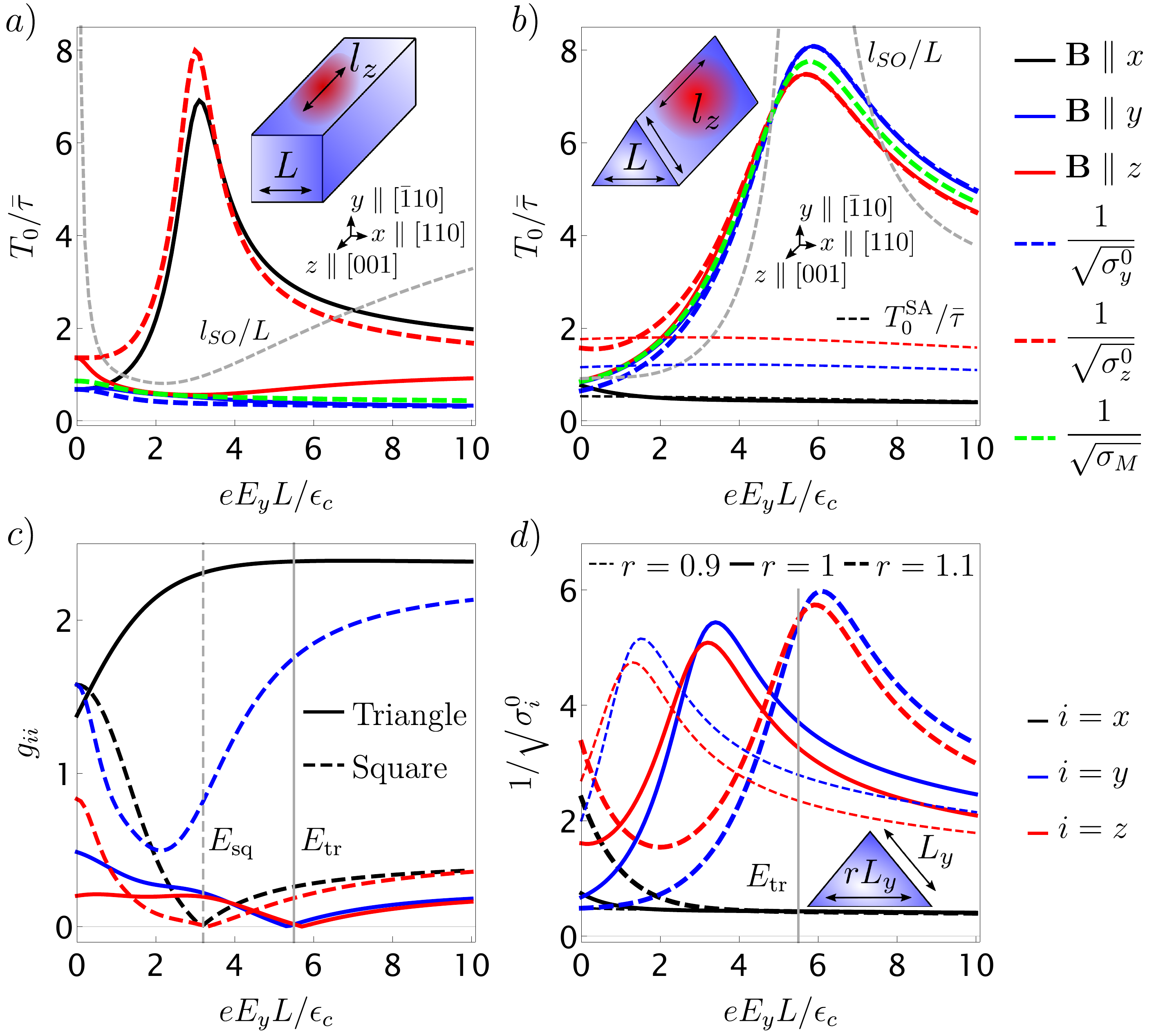}
\caption{\label{fig:E-field} Tunability of the hyperfine interactions.
In~a) and~b), we show $T_0$  against $E_y$ in Si wires grown along the DRA with square and equilateral triangular cross-sections, respectively. 
Solid (dashed) lines represent results that include (neglect) DRSOI, see Eq.~\eqref{eq:sigma_SOI}; $l_{SO}/L$ is shown in dashed gray lines.
In~b) we include the decay times $T_0^\text{SA}$ of qubits in SA wires (thin dashed lines).
In~c), we show the $g$-factors in triangular and square wire qubits. Black, blue, red lines correspond to $\textbf{B}$ aligned to $x,y,z$ direction, respectively, and we use $l_z=L$.
In~d), we include high energy states in triangular wires and compare cross-sections having different aspect ratios $r$ and the same area $\mathcal{A}_\rho=\sqrt{3}L^2/4$. Here, $L=30$~nm. 
}
\end{figure}

The presence of sweet spots for certain $L_x/L_y$ suggests that the anisotropy of the hyperfine interactions could also be externally controlled by an electric field $E_y$, which compresses the wavefunction to the upper boundary of the wire.
In Fig.~\ref{fig:E-field}a), we study a square DRA wire with side $L$, and we show the effects of the $E_y$ modulation of the wavefunction width on $T_0^{x,y,z}$ by solid black, dashed blue, and dashed red lines, respectively. As anticipated, we observe the appearance of working points at  $E_\text{sq}\approx 3.1 \epsilon_c/eL$ where hyperfine  noise is suppressed by $\sim 20$ times when $\textbf{B}\perp y$, as seen by comparing the blue curves to the black and dashed red ones in Fig.~\ref{fig:E-field}a). 
However, $E_y$ also produces a large DRSOI~\cite{DRkloeffel3,bosco2021squeezed}, that does not affect $T_0^x$ but strongly reduces $T_0^{z}$ at $E_\text{sq}$, yielding $T_0^{z}\approx T_0^y\approx \bar{\tau}/\sqrt{\sigma_M}\ll T_0^x$, see Eq.~\eqref{eq:sigma_SOI} and the solid lines in Fig.~\ref{fig:E-field}a). As a consequence, in this system the hyperfine sweet spot remains only when $\textbf{B}\parallel x$. 
We note that this field direction is useful to store information, but it is incompatible with EDSR, that requires that the Zeeman and spin-orbit fields are perpendicular to each other~\cite{PhysRevB.74.165319}. 

Strikingly, this issue is resolved in Si DRA FinFETs with triangular cross-section~\cite{bosco2020hole,geyer2020silicon,kuhlmann2018ambipolar,camenzind2021spin}, where the hyperfine sweet spots appear when $\textbf{B}\perp x$.
In fact, as shown in Fig.~\ref{fig:E-field}b), the field directions minimizing the hyperfine noise depend on the shape of the cross-section, and in remarkable contrast to square wires, in triangular fins $T_0^{y,z}\gtrsim 18 T_0^x$ at $E_\text{tr}\approx 5.9 \epsilon_c/eL$. 
Even more surprisingly, close to $E_\text{tr}$ the DRSOI is also switched off~\cite{bosco2020hole}, resulting in highly coherent qubits that are largely unaffected by both charge and hyperfine noise.
To drive these qubits, it suffices to use experimentally less demanding all-electric protocols that switch on the DRSOI by tuning $E_y$.

In Fig.~\ref{fig:E-field}b), we also compare the hyperfine noise in DRA FinFETs and in state-of-the-art devices~\cite{geyer2020silicon,kuhlmann2018ambipolar,camenzind2021spin}, where the fins are grown along the standard axes (SA) $z\parallel[110]$ and $y\parallel[100]$.
In this case, we estimate $T_0^\text{SA}\approx 0.2-0.6$~$\mu$s depending on the direction of $\textbf{B}$, in reasonable agreement with experiments~\cite{camenzind2021spin}.
At the sweet spot DRA fins yield  $\text{max}(T_0^\text{DRA})\approx 5 \text{max}(T_0^\text{SA}) $.
We emphasize that because $\bar{\tau}\propto 1/\sqrt{\nu}$ [see Eq.~\eqref{eq:time_tau}], the hyperfine noise in DRA wires is comparable to the noise in isotopically purified SA wires, where $\nu_\text{Si}^\text{iso}= 800$~ppm~\cite{veldhorst2014addressable,Yonedaquantumdotspinqubit2018} and the decay time $\text{max}(T_0^\text{SA}) $  is $7.5$ times longer than in natural Si.

We note that the position of the hyperfine sweet spots coincides with a minimal value of the $g$-factor and with a small Zeeman energy, see Fig.~\ref{fig:E-field}c). However, the hyperfine noise is suppressed in a broader range of $E_y$ in the vicinity of $E_\text{sq}$ and $E_\text{tr}$, where the $g$-factor is sizeable. 
We find also that the hyperfine sweet spots persist when high energy bands are considered~\cite{SM-hyper} and are robust against small variations of the aspect ratio $r$ of the triangular cross-section. In these cases, as shown in Fig.~\ref{fig:E-field}d), the sweet spots are shifted to different values of $E_y$.
Interestingly, when $r>1$, we observe a cross-over between a regime  where  $T_0^x\ll T_0^y$ as in equilateral triangles to a regime where  $T_0^x\gg T_0^y$ as in square wires.

\paragraph*{Hyperfine noise during qubit operations.}

\begin{figure}[t]
\centering
\includegraphics[width=0.5\textwidth]{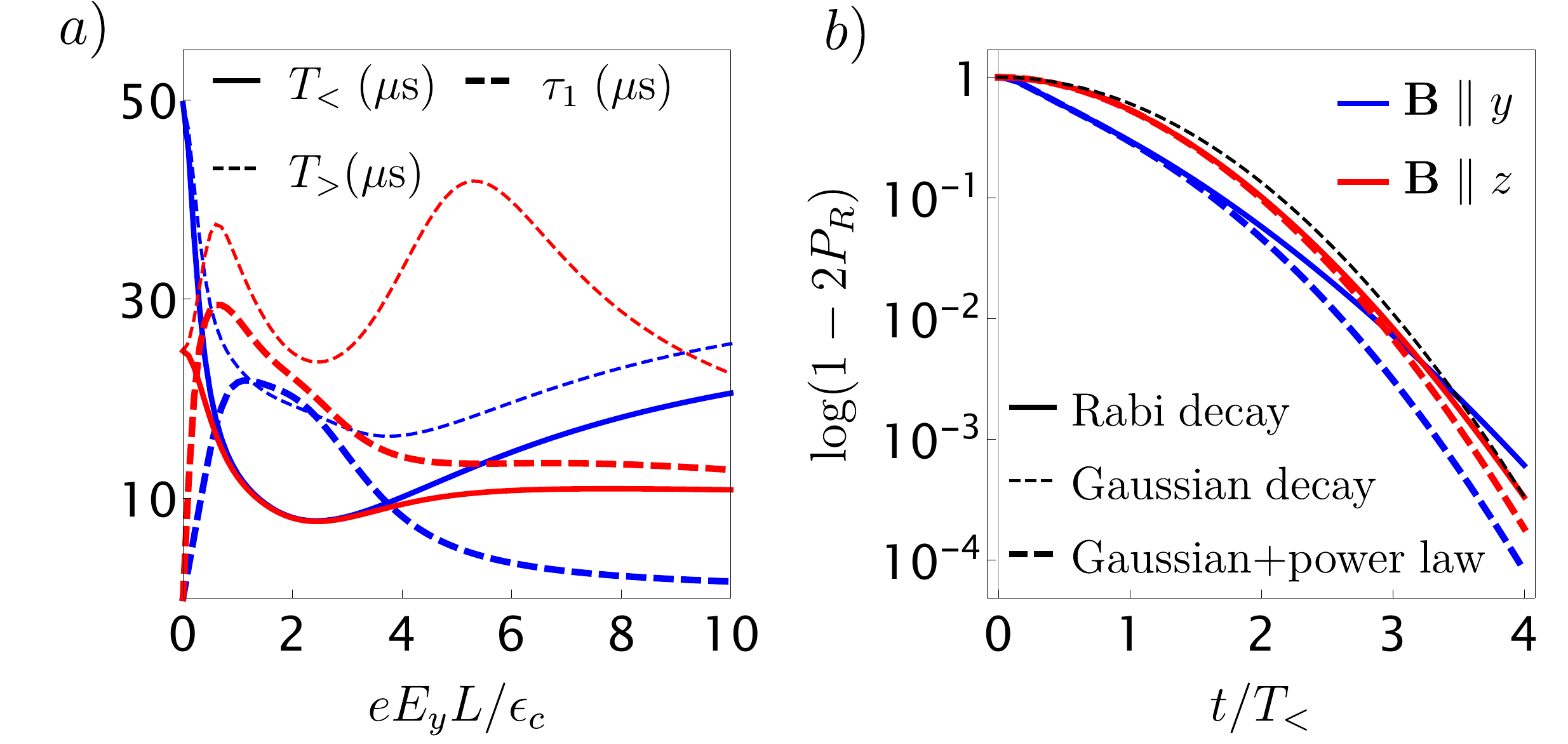}
\caption{\label{fig:Rabi} Spin decay during qubit operations. In~a) we show the relevant decay times when $\textbf{B}\parallel y$ (blue) and $\textbf{B}\parallel z$ (red) of a qubit in a Si square wire grown along the DRA. Here, $l_z=L$, $\omega_B/2\pi=3$~GHz, $N=10^4$ and $d_0=0.02 l_z$, resulting in $\omega_R^\text{max}/2\pi\approx 75$~MHz.   
In~b) we consider $E_y=6\epsilon_c/eL$ and study the spin decay of the Rabi oscillations and its deviation from purely Gaussian and power-law scaling. }
\end{figure}

The DRSOI enables fast Rabi oscillations and qubit operations via EDSR~\cite{PhysRevB.74.165319}. Neglecting small hyperfine-induced EDSR terms~\cite{PhysRevLett.99.246601,PhysRevB.78.195302}, this effect originates from an ac electric field applied along the wire, that shifts the dot from its static position by $d(t)=d_0\sin(\omega_D t)$.
To obtain the fastest  oscillations, we consider here a Zeeman field $\pmb{b}=\hbar \omega_B (0, \sin(\theta_B),\cos(\theta_B))$, perpendicular to the DRSOI, and we work at resonance $\omega_B=\omega_D$.
In a frame moving with the dot,  $\pmb{h}$ becomes time-dependent $\pmb{h}(t)$ because  $\Gamma(\textbf{r}_k)\rightarrow \Gamma(\textbf{r}_k+d(t)\textbf{e}_z)$, see Eq.~\eqref{eq:gamma}, and the spin dynamic changes drastically~\cite{PhysRevLett.116.066806}. Moving to a frame rotating with frequency $\omega_D$ around $\pmb{b}$, and in the rotating wave approximation, only a few Fourier components $\pmb{h}^m=\omega_D\int_0^{2\pi/\omega_D} dt e^{i m \omega_D t} \pmb{h}(t)/2\pi$ contribute~\cite{PhysRevLett.116.066806}, and 
\begin{equation}
\label{eq:Rabi-hamilt}
H_R\approx \frac{\hbar\omega_R+h_\perp^1}{2}\sigma_x+\frac{h^0_\parallel}{2}\sigma_z \ .
\end{equation}
The Rabi frequency is $\omega_R=\omega_D d_0/l_{SO}$,  $h_\parallel^0=h_y^0\sin(\theta_B)+h_z^0\cos(\theta_B)$, and $h_\perp^1=h_y^1\cos(\theta_B)-h_z^1\sin(\theta_B)$. A generalization of Eq.~\eqref{eq:Rabi-hamilt} valid also off-resonance is given in~\cite{SM-hyper}. 

In hole dots, $l_{SO}\sim 10$~nm~\cite{froning2020ultrafast,camenzind2021spin} is rather short and small driving amplitudes $d_0/l_z\ll 1$ suffice for fast qubit manipulation. In this case,  the covariance of $\pmb{h}$ is
\begin{equation}
\label{eq:sigma_rabi}
{\Sigma}^R= \frac{\hbar^2}{\bar{\tau}^2}\left(
\begin{array}{ccc}
  \frac{d_0^2}{4l_z^2}\left[\sigma_\perp+\frac{4l_z^2}{l_{SO}^2}\sigma_M\right] & 0 & -\frac{d_0 }{l_{SO}}\sigma_M  \\
 0  & 0  & 0 \\
 -\frac{d_0}{l_{SO}}\sigma_M  & 0 & \sigma_{\parallel}  \\
\end{array}
\right) \ ,
\end{equation}
where $\sigma_{\perp,\parallel}=\sigma_{y,z}\mp(\sigma_{y}-\sigma_z)\sin(\theta_B)^2$ are the components of the hyperfine noise parallel and perpendicular to $\pmb{b}$, respectively [see Eq.~\eqref{eq:sigma_SOI}].
From Eqs.~\eqref{eq:Rabi-hamilt} and~\eqref{eq:sigma_rabi}, one expects a Gaussian and a power-law decay determined by $\Sigma_{11}^R$ and $\Sigma_{33}^R$, respectively. We emphasize that because $\Sigma_{11}^R$ is reduced by the small parameter $d_0^2/l_z^2$, the Gaussian timescale is enlarged by $l_z/d_0$, thus enhancing the coherence of qubit operations~\cite{PhysRevLett.116.066806}.
Also, the power-law timescale  $\tau_1=\omega_R \bar{\tau}^2/\sigma_\parallel\propto d_0$ is now comparable to the Gaussian timescale even for rather isotropic hyperfine interactions, in striking contrast to idle qubits.

More precisely, the spin-flip probability averaged over a Gaussian distributed field $\pmb{h}$ with covariance $\Sigma^R$ is~\cite{SM-hyper}
\begin{equation}
\label{eq:P_Rabi}
P_R(t)\approx \frac{1}{2}-\frac{1}{2}\frac{e^{-\frac{t^2}{2T_R^2(t)}}\cos[\omega_R t+ \phi_R(t)/2]}{\sqrt[4]{1+t^2/\tau_1^2}}  \ ,
\end{equation}
with $\phi_R(t)=\arctan(t/\tau_1)-\omega_R  t^3(\Sigma^R_{13}/\Sigma^R_{33})^2/(t^2+\tau_1^2)$.
Surprisingly, we observe that the strong DRSOI not only quantitatively renormalizes ${\Sigma}^R_{ii}$, but also introduces off-diagonal elements $\Sigma_{13}^R=\Sigma_{31}^R$ that alter the spin dynamics qualitatively. In particular, they result in a time-dependent Gaussian timescale 
\begin{equation}
\frac{1}{T_R^2(t)}=\frac{1}{T_<^2}-\frac{1}{\bar{\tau}^2}\frac{d_0^2}{l_{SO}^2}\frac{\sigma_M^2}{\sigma_\parallel} \frac{t^2}{t^2+\tau_1^2}  \ ,
\end{equation}
that interpolates between the short time $T_<=\hbar/\sqrt{\Sigma^R_{11}}$, when $t\ll \tau_1$, to the longer time $T_>=\left(T_<^{-2}-{\sigma_M^2d_0^2}/{\bar{\tau}^2l_{SO}^2\sigma_\parallel}\right)^{-{1}/{2}}>T_<$, when $t\gg \tau_1$.

A comparison between different timescales in a typical Rabi experiment is shown in Fig.~\ref{fig:Rabi}a) and we note that in this case, the decay times are of tens of microseconds, much longer than in idle qubits. Rabi oscillations with similarly high coherence were recently observed in hole Si FinFETs~\cite{camenzind2021spin}.
We also predict that at the large values of DRSOI achieved in current experiments~\cite{froning2020ultrafast,camenzind2021spin,watzinger2018germanium,wang2020ultrafast} the interplay between the different decay times will yield measurable effects, see Fig.~\ref{fig:Rabi}b), further decreasing the effect of hyperfine noise during qubit operations.\\

In conclusion, we studied the hyperfine interactions of a hole spin qubit in elongated quantum dots and we showed that they can be tuned over a wide range of parameters by device design and by external electric fields.
In certain devices, this tunability enables sweet spots where the hyperfine noise is strongly reduced and becomes comparable to isotopically purified materials.
Remarkably, in Si FinFETs charge and hyperfine noises are both suppressed at these sweet spots, pushing this architecture towards new coherence standards.
Combined with the high speed and fidelity of operations, these highly coherent qubits can be reliable building blocks for scalable quantum computers.\\

We  thank Bence Het{\'e}nyi for useful discussions.
This work was supported by the Swiss National Science Foundation and NCCR SPIN.

%


\pagebreak

\onecolumngrid

\begin{center}
  \textbf{\large Supplemental Material of  \\
Fully tunable hyperfine interactions of hole spin qubits in Si and Ge quantum dots}\\[.2cm]
  Stefano Bosco and Daniel Loss\\[.1cm]
  {\itshape Department of Physics, University of Basel, Klingelbergstrasse 82, 4056 Basel, Switzerland}\\
\end{center}

\setcounter{equation}{0}
\setcounter{figure}{0}
\setcounter{table}{0}
\setcounter{section}{0}

\renewcommand{\theequation}{S\arabic{equation}}
\renewcommand{\thefigure}{S\arabic{figure}}
\renewcommand{\thesection}{S\arabic{section}}
\renewcommand{\bibnumfmt}[1]{[S#1]}
\renewcommand{\citenumfont}[1]{S#1}

\section{Hyperfine interactions in the valence band}
\label{sec:valence}
\subsection{Effective qubit Hamiltonian}
Here, we derive the general covariance matrix of the Overhauser field $\pmb{h}$ used in the main text. 
We begin from the envelope function hyperfine Hamiltonian $H_\text{HF}= \sum_k \delta(\textbf{r}-\textbf{r}_k) \textbf{h}^k\cdot \textbf{I}^k/n_0$ in a bulk semiconductor of nuclear density $n_0$, where 
\begin{subequations}
\begin{align}
{h}^k_i&=\frac{\hbar\gamma_k g \mu_B \mu_0}{4\pi}\sum_{\nu\nu'}\int_\text{U.C.} d\textbf{r}\Psi_\nu^\dagger(\textbf{r})\left[\frac{L_i(\textbf{r}-\textbf{r}_k)}{|\textbf{r}-\textbf{r}_k|^3}\sigma_0+\sum_j{T}_{ij}(\textbf{r}-\textbf{r}_k)\sigma_j \right]\Psi_{\nu'}(\textbf{r}) \ , \\
{T}_{ij}(\textbf{r})&=\frac{4\pi}{3}\delta(\textbf{r})\delta_{ij}+\frac{3 r_i r_j -|\textbf{r}|^2}{2|\textbf{r}|^5} \ , \ L_{i}(\textbf{r})=-i \textbf{r}\times\nabla\ \ , \ \ \text{with} \ \   i,j=x,y,z,
\end{align}
\end{subequations}
include contact and dipolar hyperfine interactions~\cite{1PhysRevB.78.155329,1PhysRevB.101.115302,1PhysRevB.100.085305}. Here, the integration is taken over a unit cell, $\mu_0$ is the permittivity of the medium, $\mu_B$ is the Bohr magneton and $k$ labels  nuclear defects with spin $\textbf{I}^k$ and gyromagnetic ratio $\gamma_k$ at position $\textbf{r}_k$; $g=2$ is the bare g factor.
The vector of operators $\textbf{h}^k=({h}^k_x,{h}^k_y,{h}^k_z)$ has the units of energy and acts on the space of the heavy (HH), light (LH), and split-off (SOH) holes~\cite{1WinklerSpinOrbitCoupling2003} labelled by $\nu$ and $\nu'$. 
Following Ref.~\cite{1PhysRevB.101.115302}, we decompose the wavefunctions close to the nucleus in terms of spherical harmonics $Y_{ml}(\theta,\varphi)$ as $\Psi_\nu(\textbf{r})=\sum_{ml\rho}R_{ml\rho}^\nu(r)Y_{ml}(\theta,\varphi)$ and performing the angular integration, we obtain
\begin{equation}
\label{eq:HF-matrix}
\textbf{h}^k=\frac{A_k}{2}\left(
\begin{array}{cc}
\textbf{J} & 15\textbf{T}^\dagger/16 \\
15\textbf{T}/16 & 5\pmb{\sigma}/2
\end{array}
\right) \ , \ \ \text{with} \ \ A_k=\frac{4\hbar\gamma_k g \mu_B\mu_0}{15\pi}\int \frac{dr}{r} |R_{110}(r)|^2 \ .
\end{equation}
We introduce the vectors of spin 3/2 matrices $\textbf{J}=(J_x,J_y,J_z)$, and of Pauli matrices $\pmb{\sigma}=(\sigma_x,\sigma_y,\sigma_z)$.
The components of the vector of $4\times 2$ matrices $\textbf{T}=(T_x,T_y,T_z)$ are explicitly given, for example, in Appendix C of Ref.~\cite{1WinklerSpinOrbitCoupling2003}.
We only include here $p$-type wavefunctions and neglect small corrections due to other orbitals~\cite{1PhysRevB.101.115302,1PhysRevB.100.085305}. In the main text, we consider $A_k=2A_\parallel/3$ and use the values of $A_\parallel$ estimated in Refs.~\cite{1PhysRevB.101.115302} and~\cite{1philippopoulos2020hyperfine}, for Si and Ge, respectively. In this analysis, we neglect small anisotropies $\propto J^3_i$ and long-range contributions of the dipolar hyperfine interactions~\cite{1PhysRevB.78.155329}; we also consider weak magnetic fields ($|\textbf{B}|\lesssim 1$~T) and neglect orbital effects of the magnetic field~\cite{1PhysRevResearch.3.013081}.  

To find an effective theory of a hole quantum dot, we project $H_\text{HF}$ onto the ground state subspace $\Psi_{\uparrow\downarrow} $ of the Luttinger-Kohn (LK) Hamiltonian~\cite{1WinklerSpinOrbitCoupling2003} including an electrostatic confinement  potential, $H=H_\text{LK}+V_\text{C}-e \textbf{E}\cdot \textbf{r}$.
When the dot is elongated in the $z$-direction and restricting our analysis to wires grown along a crystallographic axis, e.g. $z\parallel [001]$, $H$ reads in the basis $\left(
\left|\text{HH}\uparrow\right\rangle,\left|\text{LH}\downarrow,\right\rangle,\left|\text{LH},\downarrow\right\rangle,\left|\text{HH}\downarrow\right\rangle,\left|\text{LH}\uparrow,\right\rangle,\left|\text{LH},\uparrow\right\rangle
\right)$
\begin{subequations}
\label{eq:H-hamiltonian}
\begin{align}
\label{eq:H-hamiltonian-a}
H&=\left(
\begin{array}{cc}
H_0 & 0 \\
0 &\mathcal{T}H_0\mathcal{T}^{-1}
\end{array}
\right) +
p_z
\left(\begin{array}{cc}
0 & H_1 \\
H_1^\dagger & 0
\end{array}
\right) +
p_z^2
\left(\begin{array}{cc}
H_2 & 0  \\
 0  & \mathcal{T}H_2\mathcal{T}^{-1}
\end{array}
\right) \ ,  \\
H_0&=\frac{1}{2m}\left(
\begin{array}{ccc}
 (\gamma_1+\gamma_2) p_+p_- & \sqrt{3} \left(\gamma _- p_+^2-\gamma _+ p_-^2\right)  & -\sqrt{6}  \left(\gamma _- p_+^2-\gamma _+ p_-^2\right) \\
 \sqrt{3} \left(\gamma _- p_-^2-\gamma _+ p_+^2\right)  &(\gamma_1-\gamma_2) p_+p_- & -\sqrt{2} \gamma _2 p_+ p_- \\
 -\sqrt{6} \left(\gamma _- p_-^2-\gamma _+ p_+^2\right)   & -\sqrt{2} \gamma _2 p_+ p_- & \gamma_1  p_+p_-
  \end{array}
\right)+V_\text{C}-e \textbf{E}\cdot \textbf{r} \ , \\
H_1&=\frac{\sqrt{3}\gamma_3}{2m}\left(
\begin{array}{ccc}
 0 & -2p_- & \sqrt{2}p_-  \\
 2p_- & 0 & -\sqrt{6} p_+  \\
 \sqrt{2}p_- & -\sqrt{6} p_+  & 0
  \end{array}
\right) \ , \\
H_2&=\frac{1}{2m}\text{diag}\left(\gamma_1-2\gamma_2,\gamma_1+2\gamma_2,\gamma_1\right) \ .
\end{align}
\end{subequations}
Here, the spin quantization axis is aligned to the wire direction ($z$-direction), $p_z=-i\hbar \partial_z$ is the canonical momentum along the wire, and we introduce $\gamma_\pm=(\gamma_3\pm\gamma_2)/2$ and $p_\pm=p_{[100]}\pm i p_{[010]}$, where $p_{[100]}$ and $p_{[010]}$ are the momenta aligned to the [100] and [010] crystallographic axes, respectively. To account for different growth directions we rotate the Hamiltonian as described in Ref.~\cite{1DRkloeffel3}. The time-reversal operator $\mathcal{T}$ flips the spin, complex conjugates the wavefunctions, and changes sign of the SOHs. Our analysis can be generalized to account for strain by including the Bir-Pikus Hamiltonian~\cite{1bir1974symmetry} in $H_0$.

An effective wire theory can be derived by starting from the ground state eigenstates of $H_0$
\begin{subequations}
\begin{align}
\label{eq:WF}
\psi_\uparrow(\pmb{\rho}) &=\psi_H(\pmb{\rho})\left|\text{HH}\uparrow\right\rangle+ \psi_L(\pmb{\rho})\left|\text{LH}\downarrow\right\rangle+ \psi_S(\pmb{\rho})\left|\text{SOH}\downarrow\right\rangle \ , \\
\psi_\downarrow(\pmb{\rho}) &=\psi_H^*(\pmb{\rho})\left|\text{HH}\downarrow\right\rangle+ \psi_L^*(\pmb{\rho})\left|\text{LH}\uparrow\right\rangle- \psi_S^*(\pmb{\rho})\left|\text{SOH}\uparrow\right\rangle \ .
\end{align}
\end{subequations}
Here, $\psi_{H,L,S}$ are spin-resolved wavefunctions in the cross-section of the wire; they depend on confinement potential, on growth direction, on $\textbf{E}$, and  on strain if included. By using second order perturbation theory in $p_z$, we obtain the Hamiltonian
\begin{equation}
H_W=\frac{p_z^2}{2m^*}+\frac{m^*\omega_z^2 }{2}z^2+ v p_z \textbf{n}_\text{SO}\cdot \pmb{\sigma}=\frac{1}{2m^*}\left(p_z+\frac{\hbar}{l_{SO}}\textbf{n}_\text{SO}\cdot \pmb{\sigma}\right)^2+\frac{m^*\omega_z^2 }{2}z^2- \frac{\hbar^2}{2m^*l_{SO}^2} \ ,
\end{equation}
where the effective mass $m^*$, the spin-orbit velocity $v$, and the spin-orbit length $l_{SO}=\hbar/m^*v$ are computed in analogy to Ref.~\cite{1bosco2020hole}. In particular, 
\begin{equation}
\label{eq:SOI}
v=2\sqrt{3}\frac{\gamma_3}{m}\left|p_+^{HL}+\frac{1}{\sqrt{2}}\left( p_+^{HS}-\sqrt{3}p_-^{LS}\right)\right| \ , \ \text{with} \  p_\pm^{\alpha\beta}=\int d\pmb{\rho} \psi_\alpha(\pmb{\rho}) p_\pm \psi_\beta(\pmb{\rho})=- p_\pm^{\beta\alpha} \ .
\end{equation}
The unit vector $\textbf{n}_\text{SO}\cdot \textbf{e}_z=0$ defines the direction of the DRSOI and lies in the $\pmb{\rho}=(x,y)$ plane.
Without loss of generality, we choose the global phase of $\psi_\uparrow$ such that $\textbf{n}_\text{SO}=(1,0,0)$; this choice corresponds to a rotation of the coordinates around $z$ to a system with $\textbf{e}_1\parallel \textbf{n}_\text{SO}$, $\textbf{e}_3\parallel z$, and $\textbf{e}_2=\textbf{e}_3\times\textbf{e}_1$.

The eigenstates of $H_W$ are $\Psi_{\uparrow\downarrow}(\textbf{r})=\varphi(z) S \psi_{\uparrow\downarrow}(\pmb{\rho})$ with $\varphi(z)=e^{-z^2/2l_z^2}/\sqrt[4]{\pi l_z^2}$ and $S=e^{-i z \textbf{n}_\text{SO}\cdot\pmb{\sigma}/l_{SO}}$.
  The spin-dependent phase shift $S$ acts as a translation operator in momentum space and exactly removes the DRSOI.
These states approximate well the groundstates of $H$ in Eq.~\eqref{eq:H-hamiltonian} when the orbital energy $\hbar\omega_z$ is smaller than the energy gap $\Delta E$ between the groundstate and first excited states of $H_0$. In etched nanowires, this condition is typically met when the harmonic length $l_z=\sqrt{\hbar/m^*\omega_z}$ is larger than $L/\pi$, where $L$ is the  characteristic length of the cross-section. In addition, we remark that our analysis is valid in the weak magnetic interaction limit, where $\hbar\omega_z$ is larger than Zeeman and hyperfine energies, and it neglects corrections appearing at $|\textbf{B}|\gtrsim 1$~T.
  
The hyperfine interactions in the qubit space spanned by $\Psi_{\uparrow\downarrow}$ read $H_\text{HF}= \pmb{h}\cdot \pmb{\sigma}/2$, with  Overhauser field $\pmb{h}\equiv\sum_k A_k {\Gamma}(\textbf{r}_k) \textbf{I}^k/n_0$.
We introduce the matrix of local spin susceptibilities
\begin{equation}
\label{eq:gamma}
{\Gamma}(\textbf{r})=|\varphi(z)|^2R_1\left(\frac{2z}{l_{SO}}\right)\left(
\begin{array}{ccc}
  \text{Re}[\gamma_+(\pmb{\rho}) ]  &  \text{Im}[\gamma_-(\pmb{\rho}) ] & 0  \\
  -\text{Im}[\gamma_+(\pmb{\rho}) ]  & \text{Re}[\gamma_-(\pmb{\rho}) ]   & 0 \\
 0 & 0 & \gamma_z(\pmb{\rho})  \\
\end{array}
\right) \ ,
\end{equation}
and we define the wire susceptibilities as
\begin{subequations}
\label{eq:susc}
\begin{align}
\gamma_\pm(\pmb{\rho})&= \left(\psi_L(\pmb{\rho}) \pm \sqrt{3}\psi_H(\pmb{\rho})\right)\left(\psi_L(\pmb{\rho})+ \frac{5}{16}\sqrt{2}\psi_S(\pmb{\rho}) \right) -\frac{5}{8}\sqrt{2}\psi_S(\pmb{\rho})\left(\psi_L(\pmb{\rho})+2\sqrt{2}\psi_S(\pmb{\rho}) \right)  \ , \\
\gamma_z(\pmb{\rho})&= \frac{1}{2}\left(3|\psi_H(\pmb{\rho})|^2-|\psi_L(\pmb{\rho})|^2-5|\psi_S(\pmb{\rho})|^2\right)+\frac{5}{8}\sqrt{2}\text{Re}\left[\psi_S(\pmb{\rho})\psi_L^*(\pmb{\rho})\right] \ .
\end{align}
\end{subequations}
Explicitly, the counterclockwise rotation matrices $R_i(\theta)$ of an angle $\theta$ around the $i$th axis are given by
\begin{equation}
\label{eq:rot_wire}
R_1(\theta)=\left(
\begin{array}{ccc}
 1 &  0 & 0  \\
 0 & \cos(\theta)   &  -\sin(\theta) \\
 0 &  \sin(\theta) &  \cos(\theta)  \\
\end{array}
\right) \ , \ R_2(\theta)=\left(
\begin{array}{ccc}
  \cos(\theta)  &  0 & \sin(\theta) \\
  0 & 1 & 0 \\
  -\sin(\theta) &  0 & \cos(\theta)
\end{array}
\right) \ , \ R_3(\theta)=\left(
\begin{array}{ccc}
 \cos(\theta)   &  -\sin(\theta) & 0 \\
  \sin(\theta) &  \cos(\theta) & 0 \\
  0 &  0 & 1
\end{array}
\right) \ .
\end{equation}
These equations correctly recover the Ising limit~\cite{1PhysRevB.78.155329} when the state is purely HH-like ($\psi_L=\psi_S=0$) and reduce to the simpler ones reported in the main text when SOHs are neglected ($\psi_S=0$). In the latter case, the relation $ \underline{g}=4\kappa \int d\textbf{r} {\Gamma}(\textbf{r})$ between the matrix of $g$-factors and of local susceptibilities holds and the results shown here are directly applicable also to wires grown along $z\parallel[110]$ direction. 
To find the spin-resolved wavefunctions $\psi_{H,L}$ when $z\parallel[110]$, we consider that the Hamiltonian $H$ can be written as in Eq.~\eqref{eq:H-hamiltonian-a}, with~\cite{1DRkloeffel3} 
\begin{subequations}
\begin{align}
H_0&=\frac{1}{2m}\left(
\begin{array}{cc}
 \left(\gamma _1+\gamma _2\right) p_{[001]}^2+\left( \gamma _1+\frac{3 \gamma _3-\gamma _2}{2}\right) p_{[1\bar{1}0]}^2 & \sqrt{3} \left(\gamma_+ p_{[1\bar{1}0]}^2- \gamma _2 p_{[001]}^2+2 i \gamma _3 p_{[001]}p_{[1\bar{1}0]}\right) \\
\sqrt{3} \left(\gamma_+ p_{[1\bar{1}0]}^2- \gamma _2 p_{[001]}^2-2 i \gamma _3 p_{[001]}p_{[1\bar{1}0]}\right) & \left(\gamma _1-\gamma _2\right) p_{[001]}^2+\left( \gamma _1-\frac{3 \gamma _3-\gamma _2}{2}\right) p_{[1\bar{1}0]}^2 \\
\end{array}
\right)+V_\text{C}-e \textbf{E}\cdot \textbf{r} \ , \\
H_1&=\frac{\sqrt{3}}{m}\left( \gamma_3 p_{[001]}-i \gamma_2 p_{[1\bar{1}0]}\right)\left(
\begin{array}{cc}
 0 & -1 \\
 1 & 0 \\
\end{array}
\right) \ , \\
H_2&=\frac{1}{2m}\left(
\begin{array}{cc}
\gamma_1-\frac{3\gamma_3+\gamma_2}{2} & -\sqrt{3} \gamma_- \\
 -\sqrt{3} \gamma_- & \gamma_1 +\frac{3\gamma_3+\gamma_2}{2} \\
\end{array}
\right),
\end{align}
\end{subequations}
with $p_z=p_{[110]}$ and $p_{[1\bar{1}0]}=(p_{[100]}-p_{[010]})/\sqrt{2}$. For example, in standard Si FinFETs~\cite{1camenzind2021spin}, $p_y=p_{[001]}$  and $p_x=p_{[1\bar{1}0]}$. The spin-orbit velocity is now $ v=2\sqrt{3}\left|\gamma_3 p_{[001]}^{HL}+i \gamma_2 p_{[1\bar{1}0]}^{HL}\right|/m$, where $p_i^{HL}$ are defined as in Eq.~\eqref{eq:SOI}.

\subsection{Covariance matrix in idle qubits}
The Overhauser field $\pmb{h}$ is Gaussian distributed with a probability distribution
\begin{equation}
\label{eq:Gaussian-probability}
p(\pmb{h})= \frac{e^{-\pmb{h}{\Sigma}^{-1}\pmb{h}/2}}{\sqrt{8\pi^3|{\Sigma}|}} \ ,
\end{equation}
having zero mean and covariance matrix ${\Sigma}$, with determinant $|{\Sigma}|$.
Following Ref.~\cite{1PhysRevB.78.155329} and for an infinite-temperature thermal state, such that $\langle I^k_iI^{k'}_j \rangle= \delta_{kk'}\delta_{ij}I(I+1)/3$, we find
\begin{equation}
\label{eq:cov_Sigma}
 {\Sigma}=\sum_k \frac{A_k^2}{n_0^2} \langle[{\Gamma}(\textbf{r}_k) \textbf{I}^k]\otimes[{\Gamma}(\textbf{r}_k) \textbf{I}^k]\rangle\approx \frac{\hbar^2}{\bar{\tau}^2}\left(
\begin{array}{ccc}
 \sigma_1^0 & -e^{-\frac{l_z^2}{2l_{SO}^2}}\text{Im}[\mathcal{I}_2] & 0  \\
 -e^{-\frac{l_z^2}{2l_{SO}^2}}\text{Im}[\mathcal{I}_2]  & \sigma_{M}+e^{-\frac{2l_z^2}{l_{SO}^2}}(\sigma_{2}^0-\sigma_{M})   & 0 \\
 0 & 0 & \sigma_{M}+e^{-\frac{2l_z^2}{l_{SO}^2}}(\sigma_{3}^0-\sigma_{M})  \\
\end{array}
\right) \ .
\end{equation}
We use the envelope-function approximation $\sum_k\approx \nu n_0 \int d\textbf{r}$~\cite{1PhysRevB.78.155329}, with $\nu$ being the average percentage of defects and define the time $\bar{\tau}=\hbar\sqrt{3N}/|A_k|\sqrt{\nu I(I+1)}$, where $N=\sqrt{2\pi}l_z \mathcal{A}_{\rho}n_0$ is the number of atoms in the dot and $A_\rho$ is the area of the cross-section. The matrix elements are $\sigma_1^0=\mathcal{I}_1+ \text{Re}[\mathcal{I}_2]$,  $\sigma_2^0=\mathcal{I}_1- \text{Re}[\mathcal{I}_2]$, $\sigma_3^0=\mathcal{I}_0$, and $\sigma_M=(\sigma_2^0+\sigma_3^0)/2$, and they depend on the dimensionless integrals
\begin{equation}
\mathcal{I}_0=\mathcal{A}_{\rho} \int d\pmb{\rho}\gamma_z(\pmb{\rho})^2 \ , \ \
\mathcal{I}_1=\mathcal{A}_{\rho}\int d\pmb{\rho} \frac{|\gamma_+(\pmb{\rho})|^2+|\gamma_-(\pmb{\rho})|^2}{2} \ , \ \
\mathcal{I}_2=\mathcal{A}_{\rho}\int d\pmb{\rho}\frac{\gamma_+(\pmb{\rho})^2-\gamma_-(\pmb{\rho})^2}{2} \ .
\end{equation}
Explicitly, by using Eq.~\eqref{eq:susc} and at  $\psi_S=0$, these integrals reduce to
\begin{equation}
\mathcal{I}_0=\mathcal{A}_{\rho} \int d\pmb{\rho}\left(\frac{3}{2}|\psi_H|^2-\frac{1}{2}|\psi_L|^2\right)^2 \ , \ \
\mathcal{I}_1=\mathcal{A}_{\rho}\int d\pmb{\rho} \Big( |\psi_L|^4+3|\psi_L|^2|\psi_H|^2\Big) \ , \ \
\mathcal{I}_2=2\sqrt{3}\mathcal{A}_{\rho}\int d\pmb{\rho}\psi_L^3\psi_H \ . 
\end{equation}

\begin{figure}[t]
\centering
\includegraphics[width=0.95\textwidth]{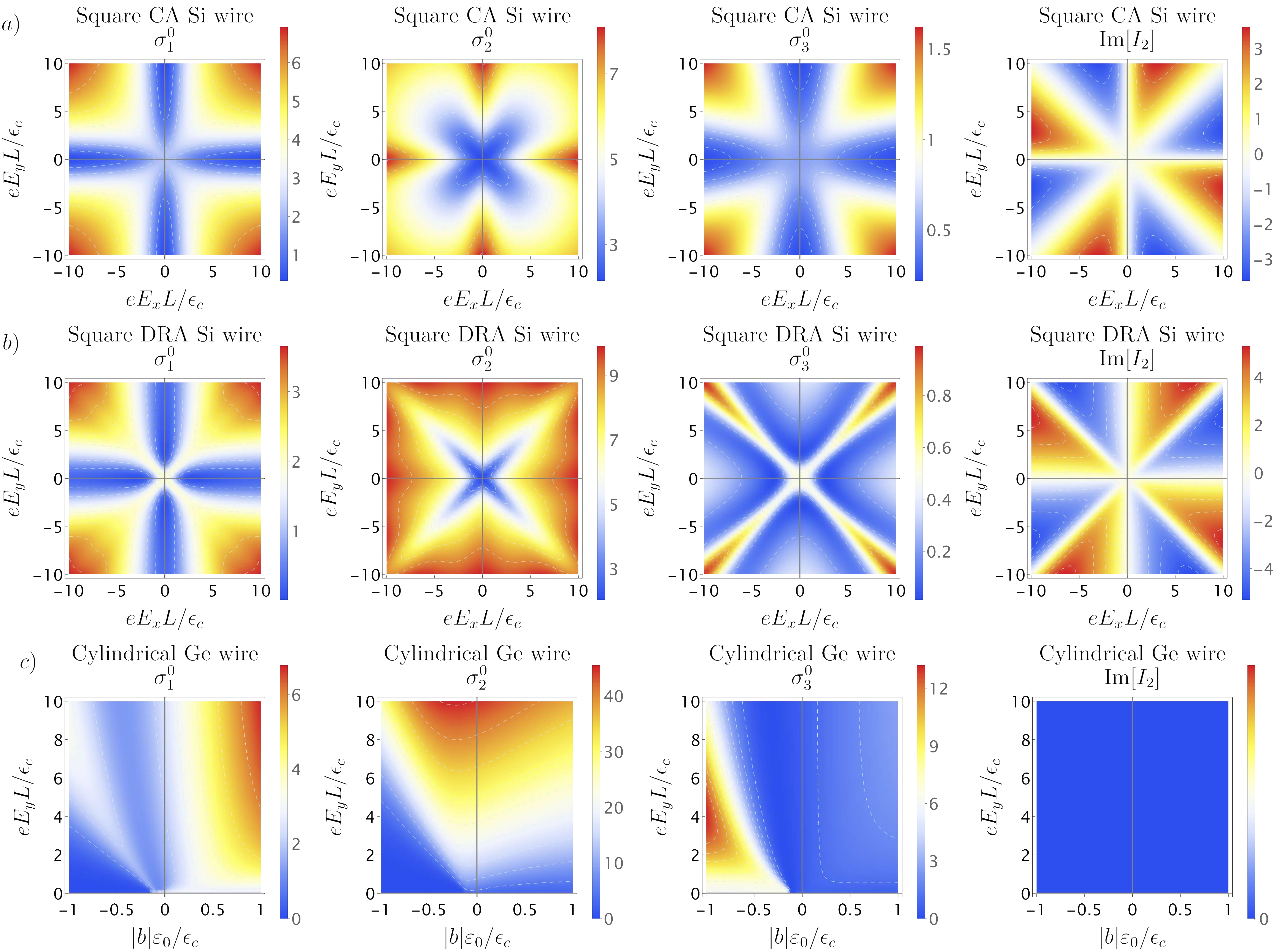}
\caption{\label{fig:covariance} Matrix elements of the covariance matrix of the hyperfine interactions in typical devices. 
In the a), b), c) rows we show the matrix elements of ${\Sigma}$ in  square Si CA wires,  square Si DRA wires, and in cylindrical Ge wires, respectively. For Ge, we use the isotropic Luttinger-Kohn Hamiltonian and thus the results are independent of the growth direction.
We show the dependence of these parameters on an electric field with a general direction and also we analyze the effect of strain, parametrized by the Bir-Pikus energy $|b|\varepsilon_0$. Here, $\epsilon_c=\hbar^2\pi^2\gamma_1/mL^2$ is the confinement energy  and $L$ is the side (radius) of the square (cylinder). 
}
\end{figure}

In Fig.~\ref{fig:covariance} we show a few experimentally relevant examples of matrix elements $\sigma_i^0$ of the covariance matrix. We neglect the SOHs in this analysis.
In Figs.~\ref{fig:covariance}a) and~\ref{fig:covariance}b), we study  square Si wires such as the ones in Ref.~\cite{1maurand2016cmos} and compare devices grown along the CA ($x\parallel[100]$ and $z\parallel[001]$) and DRA ($x\parallel[110]$ and $z\parallel[001]$), respectively. We examine the relevant hyperfine parameters as a function of the electric field components in the $x$ and $y$-direction.
The amplitude of $\sigma_{i}^0$ is highly tunable by the electric field and for certain field directions the hyperfine interactions are suppressed. The different growth directions result in different broadening and height of these resonance peaks and in particular they are sharper and higher in DRA wires.
Importantly, we also observe that the cross-terms $\Sigma_{12}\propto\text{Im}[I_2]$ are generally finite, but vanish when the electric field is aligned to a high symmetry direction, e.g. $\textbf{E}\parallel [100]$ and $\textbf{E}\parallel [110]$. These terms are neglected in the main text, where we only focus on symmetric fields, but they could modify the spin dynamics in a general case, see Sec.~\ref{sec:spin-dynamics}.   

Other common setups to define spin qubits are Ge/Si core/shell wires~\cite{1froning2020ultrafast}, where the strain produced by the Si shell is known to strongly renormalize the qubit parameters~\cite{1PhysRevB.90.115419}.
These systems can be modelled by considering strained cylindrical Ge wires under the effect of an external electric field. The hyperfine coupling in these systems is shown in Fig.~\ref{fig:covariance}c).
Because in Ge $\gamma_2\approx\gamma_3$, we used the isotropic Luttinger-Kohn Hamiltonian~\cite{1WinklerSpinOrbitCoupling2003}, that is spherically symmetric and independent of growth-directions. In this approximation,  the cross-term  of the hyperfine coupling is $\Sigma_{12}\propto\text{Im}[I_2]=0$. The strain is modelled by the Bir-Pikus Hamiltonian $|b|\varepsilon_0 J_z^2$~\cite{1bir1974symmetry}, with $|b|\approx 2.16$~eV and $\varepsilon_0=\varepsilon_\parallel-\varepsilon_{zz}$~\cite{1bosco2021squeezed}.
In these systems, the strain is produced by the mismatch of lattice constants of Si and Ge and compresses the Ge wire, resulting  in a positive valued $\varepsilon_0$~\cite{1PhysRevB.90.115419}. From Fig.~\ref{fig:covariance}c), we observe that a compressive strain with $\varepsilon_0>0$ only quantitatively renormalizes the hyperfine parameters, increasing the hyperfine coupling in the direction of the SOI and reducing it in the direction of the wire.\\

Finally, to analyze the spin dynamics it is convenient to rotate the spin-quantization axis to be aligned to the direction of the Zeeman energy $H_\text{Z}=\pmb{b}\cdot\pmb{\sigma}/2$. A general field is parametrized by the angles $\varphi_B$, $\theta_B$, and by the amplitude $\hbar\omega_B$ as
\begin{equation}
\label{eq:b-field}
\frac{\pmb{b}}{\hbar\omega_B}=  \Big[\sin(\varphi_B)\sin(\theta_B), \ \cos(\varphi_B)\sin(\theta_B), \ \cos(\theta_B)\Big] = R_B^T[0,0, 1] \ \ \text{with} \ \ R_B=R_1(\theta_B)R_3(\varphi_B)  \ ,
\end{equation} 
with $R_i$ being defined in Eq.~\eqref{eq:rot_wire}. The Overhauser field in the rotated system is $\pmb{h}_B=R_B\pmb{h}$ and has covariance matrix
\begin{equation}
\label{eq:covariance_rotation}
 {\Sigma}_B=R_B^T{\Sigma}R_B \ .
\end{equation}

For arbitrary magnetic  field directions ${\Sigma}_B$ has non-zero off-diagonal elements even if $\Sigma_{12}=0$. 
For this reason, in Sec.~\ref{sec:spin-dynamics}, we examine the spin dynamics in a general case when the covariance matrix is non-diagonal.

\subsection{Covariance matrix in driven qubits}

We now discuss the covariance matrix in an EDSR experiment.
In this case, the center of mass of the dot is shifted time-dependently by $d(t)=d_0\sin(\omega_D t)$ from the static position, resulting in a driving energy $\hbar\partial_t d(t)\sigma_1/l_{SO}$~\cite{1bosco2021squeezed}.
We now restrict ourselves to the analysis of magnetic field directions perpendicular to the DRSOI, i.e. $\varphi_B=0$ in Eq.~\eqref{eq:b-field}, and obtain the driven qubit Hamiltonian
\begin{equation}
H_Q=\frac{\hbar\omega_B}{2}\sigma_3+\hbar\omega_D\frac{d_0}{l_{SO}} \cos(\omega_D t) \sigma_1 +\frac{\pmb{h}_B(t)\cdot\pmb{\sigma}}{2}\ 
\end{equation}
after the rotation $R_B$ in Eq.~\eqref{eq:b-field}. The rotated Overhauser field $\pmb{h}_B(t)=R_B\pmb{h}(t)$ acquires a time-dependence because of the oscillations of the wavefunction and  ${\Gamma}(\textbf{r})\rightarrow {\Gamma}(\textbf{r}+d(t)\textbf{e}_z)$ in Eq.~\eqref{eq:gamma}.
In hole qubits experiments, small driving amplitudes $d_0/l_z\ll 1$ still enable ultrafast Rabi oscillations~\cite{1froning2020ultrafast}. 
For this reason, we decompose the Overhauser field in Taylor series $\pmb{h}(t)=\pmb{h}^0+ \pmb{h}^1d_0 \sin(\omega_D t)/l_z +\mathcal{O}(d_0^2/l_z^2) $ and we move to a frame rotating around $\textbf{e}_3$  with frequency $\omega_D$ by the unitary $U=e^{-i \sigma_3 \omega_D t/2}$, resulting in the Hamiltonian
\begin{equation}
H_{\text{RWA}}=U^\dagger H_Q U-i\hbar U^\dagger\partial_t U \approx \frac{\hbar\Delta}{2} \sigma_3+\frac{\hbar\omega_D}{2}\frac{d_0}{l_{SO}} \sigma_1+ \frac{\pmb{h}_\text{RWA}\cdot\pmb{\sigma}}{2} \ .
\end{equation}
Here, $\Delta=\omega_B-\omega_D$ is the detuning and in the last step we resorted to the rotating wave approximation (RWA) and neglected fast rotating terms.
The relevant hyperfine interactions in the RWA are 
\begin{equation}
\pmb{h}_\text{RWA}=\left[\frac{d_0}{2l_z}\left(h_2^1\cos(\theta_B)-h_3^1\sin(\theta_B)\right) , \  -\frac{d_0}{2l_z}h_1^1, \ h_3^0\cos(\theta_B)+h_2^0\sin(\theta_B) \right]+\mathcal{O}\left(\frac{d_0^2}{l_z^2}\right) \ ,
\end{equation}
and have still zero mean and covariance matrix 
\begin{equation}
\label{eq:covariance_RWA}
{\Sigma}_\text{RWA}=\frac{\hbar^2}{\bar{\tau}^2}\left(
\begin{array}{ccc}
\frac{d_0^2}{4l_z^2}\left[\sigma_\perp+\frac{4l_z^2}{l_{SO}^2}\sigma_M\right] &  -\frac{d_0^2(l_z^2+l_{SO}^2)}{4l_{SO}^2l_z^2}\sigma_{12}\cos(\theta_B) & -\frac{d_0}{l_{SO}}\sigma_M  \\ 
 -\frac{d_0^2(l_z^2+l_{SO}^2)}{4l_{SO}^2l_z^2}\sigma_{12}\cos(\theta_B)  & \frac{d_0^2}{4l_{z}^2}\sigma_{11}   & \frac{d_0}{2l_{SO}}\sigma_{12}\cos(\theta_B) \\
 -\frac{d_0}{l_{SO}}\sigma_M &  \frac{d_0}{2l_{SO}}\sigma_{12}\cos(\theta_B) & \sigma_{\parallel}  \\
\end{array}
\right) \ .
\end{equation}
Here, $\sigma_{\parallel}=\sigma_{33}+(\sigma_{22}-\sigma_{33})\sin(\theta_B)^2$,  $\sigma_{\perp}=\sigma_{22}-(\sigma_{22}-\sigma_{33})\sin(\theta_B)^2$, and we defined the dimensionless elements $\sigma_{ij}=\bar{\tau}^2\Sigma_{ij}/\hbar^2$ of the matrix ${\Sigma}$ in Eq.~\eqref{eq:cov_Sigma}.
Only the zeroth order term of the hyperfine interactions contributes in the direction parallel to the Zeeman energy, while in the transverse directions the terms linear in $d_0$ are relevant. Because of the $\pi/2$ phase shift between the driving and the hyperfine interactions, there is an extra $\pi/2$ rotation of the  vector $\pmb{h}^1$ around $\textbf{e}_3$, and in the RWA the interactions parallel to the driving are the original interactions in the $\textbf{e}_2$ direction. 
The same results can be obtained by first decomposing $\pmb{h}(t)$ in Fourier series with coefficients $\pmb{h}^m=\omega_D\int_0^{2\pi/\omega_D} dt e^{i m \omega_D t} \pmb{h}(t)/2\pi$ and then Taylor expanding the results in $d_0/l_z$. This latter procedure also can be generalized to large driving amplitudes~\cite{1PhysRevLett.116.066806}.
For isotropic hyperfine interactions and small SOI,  Eq.~\eqref{eq:covariance_RWA} agrees with the weak driving limit in~\cite{1PhysRevLett.116.066806}. 
In the main text, we neglect the small contributions coming from the $\sigma_{11}d_0^2/4l_z^2$ terms and we study only cases where $\sigma_{12}=0$.

To account for a finite detuning, we further rotate the spin of an angle $\theta_R=-\arctan\left[d_0\omega_D/l_{SO}\Delta\right]$ by the transformation $U_\Delta=e^{i \sigma_2 \theta_R/2}$. 
The resulting Hamiltonian is 
\begin{equation}
H_R=U_\Delta^\dagger H_\text{RWA} U_\Delta= \frac{\hbar \omega_R}{2}\sigma_3+\frac{\pmb{h}_R \cdot \pmb{\sigma}}{2} \ ,
\end{equation}
with Rabi frequency $\omega_R=\sqrt{\omega_D^2 d_0^2/l_{SO}^2+\Delta^2}$ and hyperfine interactions $\pmb{h}_R=R_2(\theta_R)\pmb{h}_\text{RWA}$, see Eq.~\eqref{eq:rot_wire}.
In this case, the covariance matrix in Eq.~\eqref{eq:covariance_RWA} transforms as ${\Sigma}_R=R_2(\theta_R)^T{\Sigma}_\text{RWA}R_2(\theta_R)$.

\section{Spin dynamics with a general covariance matrix}
\label{sec:spin-dynamics}

\begin{figure}[t]
\centering
\includegraphics[width=0.9\textwidth]{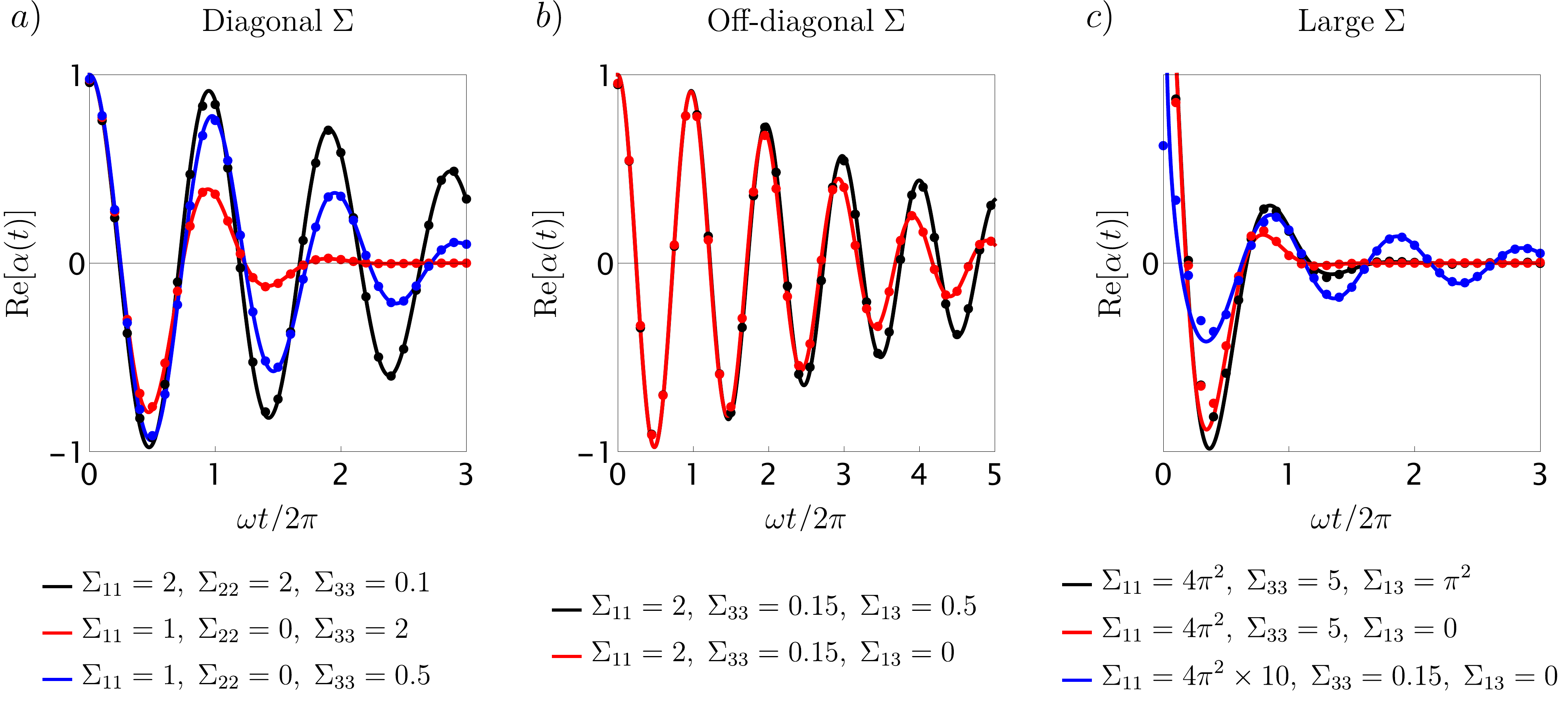}
\caption{\label{fig:variations} Approximate spin dynamics.
We compare the numerical solutions (dots) of the integral in Eq.~\eqref{eq:alpha} with the analytical approximation (lines) in Eq.~\eqref{eq:alpha-large} when the off-diagonal elements of $\Sigma$ are zero (a) and finite (b). 
The matrix elements non specified in the legends are zero and the values of $\Sigma_{ij}$ are given in units of $(\hbar \omega/2\pi)^2$.  
In c), we analyze the spin decay when the hyperfine broadening is comparable to $\hbar\omega$. Strikingly, while Eq.~\eqref{eq:alpha} fails to capture the spin dynamics at short times, it still nicely reproduces the decay at longer times. }
\end{figure}

As shown in Sec.~\ref{sec:valence}, in the cases studied here, the spin dynamics can be captured by an Hamiltonian of the form
\begin{equation}
H=\frac{\hbar \omega}{2}\sigma_3+\frac{\pmb{h}}{2}\cdot\pmb{\sigma} \ ,
\end{equation}
with $\pmb{h}$ being Gaussian distributed with probability $p(\pmb{h})$, see Eq.~\eqref{eq:Gaussian-probability}, having a non negative covariance matrix ${\Sigma}$.
We now study the average spin-flip probability $P(t)$ between the eigenstates $|+\rangle$ and $|-\rangle$ of $\sigma_1$, i.e. 
\begin{equation}
\label{eq:alpha}
P(t)=\int d\pmb{h} p(\pmb{h}) |\langle-|e^{-i H t/\hbar}|+\rangle|^2 \equiv \frac{1}{2}\text{Re}\left[\alpha(0) -\alpha(t) \right]  \ , \ \text{with} \ \alpha(t)=\int d\pmb{h} p(\pmb{h}) \frac{h_y^2+(h_z+\hbar\omega)^2}{|H|^2}e^{i {|H| t}/{\hbar}}\ ,
\end{equation}
and $|H|=\sqrt{h_x^2+h_y^2+(h_z+\hbar\omega)^2}$.
This probability describes the Rabi oscillations when $\omega=\omega_R$, and the dephasing of an idle qubit prepared in a $\sigma_x$ eigenstate when $\omega=\omega_B$.

We are not able to find a general expression for $\alpha(t)$ for arbitrary field strengths, however, progress can be made when the field splitting is large compared to the hyperfine broadening, i.e. $\hbar\omega\gg \sqrt{\Sigma_{ii}}$, with $i=(1,2,3)$. This is the usual case in the examples analyzed in the main text.
The integral $\alpha(t)$ then can be approximated as
\begin{equation}
\label{eq:alpha-large}
\alpha(t)\approx \int d\pmb{h} p(\pmb{h}) e^{i \left( \omega+\frac{h_z}{\hbar}+\frac{h_x^2+h_y^2}{2\hbar^2\omega}\right) t} = \frac{e^{i\omega t-\frac{t^2}{2}     \left[\frac{1}{T_0^2}+\frac{1}{T_+^2}\frac{it/\tau_+}{1-it/\tau_+}+\frac{1}{T_-^2}\frac{it/\tau_-}{1-it/\tau_-} \right]}}{\sqrt{(1-it/\tau_+)(1-it/\tau_-)}} \ ,
\end{equation}
where we introduce the real valued quantities
\begin{subequations} 
\begin{align}
T_0&=\frac{\hbar}{\sqrt{\Sigma_{33}}} \ , \\
\tau_{\pm}&=\frac{2\hbar^2\omega}{ \Sigma_{11}+\Sigma_{22}\pm\sqrt{\left(\Sigma_{11}-\Sigma_{22}\right)^2+4\Sigma_{12}^2}} \ , \\
\frac{1}{T_\pm^{2}}&=\pm \frac{1}{\hbar^2(\tau_+-\tau_-)}\left[  \frac{\Sigma_{13}^2\Sigma_{22}+\Sigma_{23}^2\Sigma_{11}-2\Sigma_{12}\Sigma_{13}\Sigma_{23}}{\Sigma_{11}\Sigma_{22}-\Sigma_{12}^2}\tau_\pm -\frac{\Sigma_{13}^2+\Sigma_{23}^2}{\Sigma_{11}+\Sigma_{22}}(\tau_++\tau_-)   \right] \ .
\end{align}
\end{subequations} 
When $\Sigma_{13}=\Sigma_{23}=0$, then $1/T_\pm^2=0$ and we recover the typical Gaussian and power law decay. When $\Sigma_{12}=0$, the power law times simplify as $\tau_\pm= \hbar^2\omega/\Sigma_{11,22}$ and $(1/T_+^{2}, 1/T_-^{2})=(\Sigma_{13}^2/\hbar^2\Sigma_{11},\Sigma_{23}^2/\hbar^2\Sigma_{22})$.

Explicitly, taking the real part and using $\alpha(0)\approx 1$, we find
\begin{subequations}
\begin{align}
P(t)&\approx \frac{1}{2}\left(1-\frac{e^{-\frac{t^2}{2 T_G(t)^2}} \cos\left[\omega t +\phi(t)/2\right]}{\sqrt[4]{(1+t^2/\tau_+^2)(1+t^2/\tau_-^2)}} \right)\ , \\
\frac{1}{T_G(t)^2}&=\frac{1}{T_0^2}-\frac{1}{T_+^2}\frac{t^2}{t^2+\tau_+^2}-\frac{1}{T_-^2}\frac{t^2}{t^2+\tau_-^2}  \ , \\
\phi(t)&= \arctan(t/\tau_+)+\arctan(t/\tau_-)- \frac{1}{T_+^2}\frac{\tau_+ t^3}{t^2+\tau_+^2}-\frac{1}{T_-^2}\frac{\tau_- t^3}{t^2+\tau_-^2} \ .
\end{align}
\end{subequations}

In Fig.~\ref{fig:variations}, we compare the approximation in Eq.~\eqref{eq:alpha-large} (lines) and the values of $\text{Re}[\alpha(t)]$ computed numerically from the exact Eq.~\eqref{eq:alpha} (dots). In Fig.~\ref{fig:variations}a) we observe an excellent agreement even at rather large values of $\Sigma_{ii}\sim 2 (\hbar\omega/2\pi)^2$, which justifies the use of this approximation in the regimes considered in the main text. 
In Fig.~\ref{fig:variations}b), we show the enhancement of the Gaussian decay time by the cross-term $\Sigma_{13}$. 
In Fig.~\ref{fig:variations}c), we explore the limits of validity of the approximation. At comparable sizes of $\Sigma_{ii}\sim\hbar^2\omega^2$, Eq.~\eqref{eq:alpha-large} fails to describe the short time behaviour of the spin because in this case $\alpha(0)\neq 1$, but it captures reasonably well the spin decay at longer times even when~$\Sigma_{ii}>\hbar^2\omega^2$.

\end{document}